\documentclass{ws-ijmpd}
\usepackage[super,compress]{cite}
\usepackage{color}
\begin{document}

\markboth{Authors' Names}
{Instructions for Typing Manuscripts (Paper's Title)}

%
\catchline{}{}{}{}{}
%

\title{GUP corrections to black hole thermodynamics in the extended phase space approach }

\author{Eduardo L\'opez}

\address{Departamento de Matem\'aticas, Colegio de Ciencias e Ingenier\'ia, Universidad San Francisco de Quito, Quito, Ecuador}

\author{Pedro Bargue\~no}

\address{Departamento de F\'{\i}sica Aplicada, Universidad de Alicante, Campus de San Vicente del Raspeig, E-03690 Alicante, Spain\\
pedro.bargueno@ua.es}

\author{Ernesto Contreras}

\address{Departamento de F\' isica, Colegio de Ciencias e Ingenier\'ia, Universidad San Francisco de Quito, Quito, Ecuador\\
econtreras@usfq.edu.ec}
\maketitle

\begin{history}
\received{Day Month Year}
\revised{Day Month Year}
\end{history}

\begin{abstract}
In this work, we study corrections to black hole temperature and entropy in the context of the generalized uncertainty principle. In particular, we obtain corrected asymptotically anti de–Sitter black hole solutions following the extended phase space scheme in which the cosmological constant is considered as a thermodynamic pressure satisfying certain equation of state. Among all the possibilities, we consider that the cosmological pressure satisfies either a Polytropic or a Chaplygin equation of state. The physical plausibility of the solutions is studied based on the energy conditions and the associated heat capacity.
\end{abstract}

\keywords{GUP; Asymptotically AdS Black Holes; Black Hole Thermodynamics.}

\ccode{PACS numbers:}


\section{Introduction} \label{sec:introduction}

A it is well---known, General Relativity is incompatible with the Heisenberg uncertainty principle in the sense that, obtaining an accurately measurement of the position requires a large amount of energy which, eventually, could lead to the breaking--down of the space--time structure. In the past decades, several efforts have made with the aim to define a Generalized uncertainty principle (GUP) \cite{Das:2008kaa,Ali:2010yn,Bargueno:2015tea} (see also \cite{Lambiase:2022xde,Kanzi:2021jrl,Lake:2021gbu}, for recent developments) to avoid such singularities in the energy sector of GR. One interesting feature about GUP proposals it that they lead to
a model--independent way to deduce the discreetness of the space-time \cite{Ali:2009zq}. Consequently, the deformed canonical commutation relations which arise from the particular GUP under consideration, have been used to compute corrections of the energy shift of the hydrogen atom spectrum, the Lamb shift, the Landau levels, and the Scanning Tunneling Microscope (STM) \cite{Ali:2010yn,Vagenas:2017vsw,Ali:2011fa,Pedram:2011xj}. Besides, based on the Heisenberg microscope argument, is possible to find an expression for the temperature of a black hole (BH) which  deviates from the standard formula for the Hawking temperature \cite{Vagenas:2017vsw,Azizi:2022yby,Casadio:2020rsj}. Recently, with the aim to investigate both static and spherically symmetric and Kerr-Newman BHs in the context of GUP, in Ref. \cite{okc19} the authors considered a GUP--corrected BH temperature to obtain a modified Van der Waals BH.
\\

The study of BH thermodynamics can be traced back to the seminal works by Bekenstein \cite{Bekenstein:1972tm,Bekenstein:1973ur} and Hawking \cite{Bardeen:1973gs} setting the basis for the interpretation of the gravity as an emergent phenomena. In recent developments, it has been demonstrated that the role played by the cosmological constant as a pressure, leads to the realization that BH thermodynamics is much richer than previously thought. The analysis of the resulting BH thermodynamics leads to the extended phase space (EPS) approach. In this work, we follow the route developed in \cite{okc19} in order to obtain the GUP corrected solutions for Polytropic BH \cite{set15} and Chaplygin BH \cite{deb19} obtained through the EPS approach.
\\

 This work is organized as follows. In Chapter \ref{sec:EPS}, the theoretical background contains two main sections. On one hand, the section \ref{sec:GUP} is a review of the extended phase space, which introduces negative cosmological constant as thermodynamic pressure. On the other hand, the section \ref{sec:GUP} is a review of the independent result GUP and some possible physical modifications. In Chapter \ref{sec:RD}, we develop the GUP thermodynamics strategy and we calculate black hole asymptotic GUP corrections for Polytropic BH and Chaplygin BH. In Chapter \ref{sec:T}, we calculate thermodynamic quantities to determine the stability of the GUP corrected solutions. Finally, the last Chapter \ref{sec:CO} contains final comments and beyond perspectives of this work.
	
\section{Extended phase space }\label{sec:EPS}

As it is well known the basic thermodynamic quantities of a physical system have their counterpart in BH physics. For example, for a Schwarzschild BH the mass is related to the energy, the surface gravity is related to the temperature and the horizon area is associated with the entropy of the system. However, for non--vacuum solutions where pressure and volume terms are introduced \cite{kub15} the above correspondence is not suitable to describe BH thermodynamics. For example, when considering asymptotically AdS BHs, the mass must be identified with the enthalpy of system \cite{cal00, kas09} and the cosmological constant as the thermodynamic pressure, namely  \cite{kub15,kub17}. 
\begin{eqnarray}\label{pressure}
P = -\frac{\Lambda}{8 \pi} = \frac{3}{8 \pi l^{2}}.
\end{eqnarray}
The set of thermodynamic variables which satisfies the laws of BH thermodynamics is known as extended phase space. In terms of these variables, the first law of BH thermodynamics reads,
\begin{eqnarray}\label{firstlaw}
\delta M = T \delta S + V \delta P + ...,
\end{eqnarray}
where 
\begin{eqnarray}\label{volume}
V=\frac{\partial M}{\partial P}\bigg|_{S} ,
\end{eqnarray}
is the associated thermodynamic volume. 
	
To fix ideas, let us consider the metric
\begin{eqnarray}\label{le}
ds^{2}=-f dt^{2}+f^{-1}dr^{2}+r^{2}(d\theta^{2}+\sin^{2}\theta d\phi^{2}),  
\end{eqnarray}
with  
\begin{eqnarray}\label{f}
f=-\frac{2M}{r}+\frac{r^{2}}{l^{2}}-h(r,P),
\end{eqnarray}
to ensure an asymptotically AdS solution  of the Einstein field equations, which read 
\begin{eqnarray}\label{eins}
G_{\mu\nu}+\Lambda g_{\mu\nu}=8\pi T_{\mu\nu},
\end{eqnarray}
where, the energy momentum tensor is defined as
\begin{eqnarray}\label{tmunu}    
T^{\mu}_{\nu} =diag (-\varrho, p_{r}, p_{\perp}, p_{\perp}).
\end{eqnarray}
Now, by using Eqs. (\ref{le}), (\ref{eins}) and (\ref{tmunu}) we obtain
\begin{eqnarray}
\label{rh} \varrho=-p_{r}&=&\frac{1-f-rf'}{8\pi r^{2}} + P, \\
\label{pp} p_{\perp}&=&\frac{rf''+2f'}{16\pi r}-P,
\end{eqnarray}
which results in a system of two equations  for three unknowns, $\{f,\varrho, p_{\perp}\}$. Furthermore, the physical acceptability of the solution is restricted by the energy conditions, namely the null (NEC), weak (WEC), strong (SEC) and dominant (DEC) energy conditions,
\begin{eqnarray}
\quad \varrho + p_{\perp} & \geq & 0 \label{nec},\\
\varrho \geq 0, \quad & \textnormal{and} & \quad \varrho + p_{\perp} \geq 0 \label{wec},\\
2 p_{\perp} \geq 0, \quad & \textnormal{and} & \quad \varrho + p_{\perp} \geq 0 \label{sec},\\
\varrho \geq 0, \quad & \textnormal{and} & \quad \varrho \geq \lvert p_{\perp} \rvert \label{dec},
\end{eqnarray}
respectively.

Now, if we were solving the Einstein field equations in the usual way, the next step should be to propose an equation of state between the variables of the matter sector. However, we can follow an alternative protocol which consists in proposing an equation of state (EoS) for the pressure $P$ and then use the equations of BH thermodynamics to solve the system. At this point, some comments are in order. First note that from the horizon condition, $f(r_{+})=0$, we obtain
\begin{eqnarray}\label{Smar}
M=\frac{4\pi}{3} r_{+}^{3}P -\frac{r_{+}}{2}h(r_{+},P).
\end{eqnarray}
Then, for Eq. (\ref{firstlaw}), the thermodynamic volume reads 
\begin{eqnarray}\label{volu}
V=\bigg(\frac{\partial M}{\partial P} \bigg)_{S} = \frac{4\pi}{3} r_{+}^{3} - \frac{r_{+}}{2}\frac{\partial h(r_{+},P)}{\partial P}.
\end{eqnarray}
Similarly, the temperature can be expressed as
\begin{eqnarray} \label{temp}
T = \frac{f'(r_{+},P)}{4\pi} = 2r_{+}P-\frac{h(r_{+},P)}{4\pi r_{+}}-\frac{1}{4\pi}\frac{\partial h(r_{+},P)}{\partial r_{+}}. \label{temph}
\end{eqnarray}
Finally, we consider the well established Bekenstein-Hawking entropy given by
\begin{eqnarray}\label{entr}
S=\frac{A}{4}=\pi r_{+}^{2}.
\end{eqnarray}

Note that, in order to close the system of equations,
an EoS for the pressure $P$ must be provided. In the next section, we shall explore some of the equations of state that have been proposed so far to obtain asymptotically AdS BH solutions.

\subsection{Polytropic BH}\label{polysect}

In this section we will consider a polytropic EoS (see Ref. \cite{set15}, for details), namely
\begin{eqnarray}\label{EoSpoly}
P = K \rho^{1 + \frac{1}{n}} \quad ; \quad \rho = K^{-\frac{n}{n+1}} P^{\frac{n}{n+1}} = C P^{\frac{n}{n+1}},
\end{eqnarray}
where $n$ is the Polytropic index and $K$ and $C$ are constants, related as
\begin{eqnarray}
C=K^{-\frac{n}{n+1}}.
\end{eqnarray}
Additionally, it is assumed that the gas satisfies the following integrability condition
\begin{eqnarray}
\frac{\partial^{2}S}{\partial T\partial V}=
\frac{\partial^{2}S}{\partial V\partial T},
\end{eqnarray}
from where, in accordance with the first law of thermodynamics, we obtain
\begin{eqnarray}\label{flt}
S=\frac{P+\rho}{T} V .
\end{eqnarray}
In order to proceed, we will consider the following ansatz, proposed in \cite{set15}:
\begin{eqnarray}\label{anzatsfltpoly}
h(r,P) = A(r)-B(r)P+D(r)P^{\frac{1}{1+n}}.
\end{eqnarray}
Then, by using the thermodynamic relations \eqref{volu}, \eqref{temp} and \eqref{entr}, the polytropic EoS \eqref{EoSpoly}, and the first law \eqref{flt}, we obtain
\begin{eqnarray}\label{solanzfltpoly}
F_{1}(r)+F_{2}(r)P+F_{3}(r)P^{\frac{1}{1+n}}+F_{4}(r)P^{\frac{n}{1+n}}=0,
\end{eqnarray} 
where $F_{1}$, $F_{2}$, $F_{3}$ and $F_{4}$ are functions o $A(r)$, $B(r)$, $D(r)$ and their derivatives.
Now, by setting the condition $F_{4}(r) = 0$, we obtain
\begin{eqnarray}\label{solanzfltpolyB}
B(r) = - \frac{8\pi r^{2}}{3} + r B_{0}.
\end{eqnarray}
Then, the condition $F_{3}(r) = 0$ leads to
\begin{eqnarray}\label{solanzfltpolyD}
D(r) = D_{0} [(n+1) r]^{\frac{1-n}{n+1}}.
\end{eqnarray}
It is worth mentioning that the solution  \eqref{solanzfltpolyB} is consistent with the remaining equation, $F_{2}(r)$ = 0. Finally, by imposing $F_{1}(r) = 0$, the function $A(r)$ reads
\begin{eqnarray}\label{solanzfltpolyA}
A(r) = \frac{A_{0}}{r} + D_{0}C [(n+1)r]^{\frac{1-n}{n+1}}.
\end{eqnarray}
With these results at hand, the function $h(r,P)$ takes the form
\begin{eqnarray}\label{solanzfltpolyh}
h(r,P)&=& \frac{A_{0}}{r}+\frac{r^2}{l^2}-\frac{3 B_{0} r}{8 \pi  l^2}+D_{0} \bigg[ C + \bigg(\frac{3}{8 \pi l^2}\bigg)^{\frac{1}{n+1}} \bigg] (n+1)^{\frac{1-n}{n+1}} r^{\frac{1-n}{n+1}}
\end{eqnarray}
Finally, as stated in \cite{set15}, we need to impose 
$A_{0}=B_{0}=0$ and $D_{0}=\bigg\{ l^{2} \bigg[ C + \bigg(\frac{3}{8 \pi l^2}\bigg)^{\frac{1}{n+1}} \bigg] (n+1)^{\frac{1-n}{n+1}} \bigg\}^{-1}$ for $n=-\frac{1}{3}$, to have the asymptotically AdS behaviour.
	
\subsection{Chaplygin BH}\label{chapsect}

A different possibility is to consider the Chaplygin EoS and then to follow the same steps used in the polytropic case as done in Ref. \cite{deb19}. To be more precise, we start from  
\begin{eqnarray}\label{EoSchap}
P = A \rho - \frac{B}{\rho^{n}},
\end{eqnarray}
where $A$, $B$ and $n$ are constants. Note that in this case,  the extended phase space formalism remains applicable with a modification in the volume, 
\begin{eqnarray}\label{voluchap}
V=\bigg(\frac{\partial M}{\partial P} \bigg)_{S} = \frac{4\pi}{3} r_{+}^{3} - \frac{r_{+}}{2}\frac{\partial h(r_{+},\rho)}{\partial \rho} \frac{\partial \rho}{\partial P}.
\end{eqnarray}
Next, as reported in \cite{deb19},  
the ansatz for $h$ is taken as
\begin{eqnarray}\label{anztasfltchap}
h(r, \rho) = X(r) + Y(r) \rho + Z(r) \rho^{-n},
\end{eqnarray}
from where, following the same steps as in the polytropic case,
we obtain a polynomial equation of the form
\begin{eqnarray}\label{solanzfltchap}
&&F_{0} + F_{1} \rho + F_{2}(r) \rho^{-n} + F_{3}(r) \rho^{-n - 1}\nonumber\\
&&\hspace{2cm}+ F_{4}(r) \rho^{-2 n - 1} = 0
\end{eqnarray}
Setting the condition $F_{0}(r) = F_{3}(r) = 0$ leads to
\begin{eqnarray}\label{solanzfltchapX}
X(r) = \frac{X_{0}}{r}.
\end{eqnarray}
Now, from $F_{4}(r)=0$ we obtain
\begin{eqnarray}\label{solanzfltchapY}
Y(r) = \frac{8}{3} \pi  A r^2 + Y_{0} r^{\frac{2}{A}+1},
\end{eqnarray}
and from $F_{2}(r)=0$ we have
\begin{eqnarray}\label{solanzfltchapZ}
Z(r) = -\frac{8}{3} \pi B r^2 + r Z_{0}.
\end{eqnarray}
Using \eqref{solanzfltchapY} and \eqref{solanzfltchapZ} in  $F_{1}(r)=0$, results in 
\begin{eqnarray}\label{solanzfltchapA}
A = - \frac{n}{1+n}.
\end{eqnarray}
Finally, replacing the results \eqref{solanzfltchapX}, \eqref{solanzfltchapY},
\eqref{solanzfltchapZ} and
\eqref{solanzfltchapA} in Eq. \eqref{anztasfltchap}, the expression for $h(r,\rho)$ reads
\begin{eqnarray}\label{hdep}
h(r,\rho) &=& \frac{X_{0}}{r}+\frac{r^2}{l^2}+\rho Y_{0} r^{-\frac{2}{n}-1}\nonumber\\
&&-\frac{3 r Z_{0}}{8 \pi  B l^2}-\frac{n \rho  r Z_{0}}{B(n+1)},
\end{eqnarray}
where $X_{0}$, $Y_{0}$, and $Z_{0}$ that should be set  $X_{0}=Z_{0}=0$, $Y_{0}=\frac{8 \pi P}{3 \rho}$, and $n=-\frac{2}{3}$ in order to ensure the asymptotically AdS behaviour \cite{deb19}.

Before concluding this section, we would like to emphasize the importance of the extended phase space formalism as an alternative protocol in obtaining asymptotically AdS BH's. Note that, given an equation of state for the pressure $P$, we can use equations (\ref{volu}), (\ref{temp}) and (\ref{entr}) to determine the metric instead of solving Einstein field equations directly. However, the results obtained depend on how both the temperature and entropy are written in terms of the metric and coordinates so that, any modification in such quantities would lead to different results. In this work we shall assume a GUP--corrected temperature to solve Eqs. (\ref{volu}), (\ref{temp}) and (\ref{entr})
following the same steps that let us to Eqs. (\ref{solanzfltpolyh}), (\ref{hdep}).

\section{Generalized uncertainty principle} \label{sec:GUP}

\subsection{Minimal Length Uncertainty}

As it is well known, the GUP states that the fundamental commutator is modified as
\begin{eqnarray}
[x_{i},p_{j}]=i\hbar [\delta_{ij}(1+\alpha p^{2})+2\alpha p_{i}p_{j}],
\end{eqnarray}
with
\begin{eqnarray}
[x_{i},x_{j}]=0=[p_{i},p_{j}],
\end{eqnarray}
being $\alpha$ a positive constant, called the GUP parameter.

Now, we can define
\begin{eqnarray}
x_{i}&=&x_{0i}\\
p_{j}&=&p_{0j}(1+\alpha p_{0}^{2}),
\end{eqnarray}
where $[x_{0i},p_{0j}]=i\hbar \delta_{ij}$. Furthermore, $p_{0j}$ is interpreted as the momentum at low energy scale which is represented by $p_{0j}=-i\hbar \frac{\partial}{\partial x_{0j}}$, while $p_{j}$ is considered as the momentum a high energy scales. Now the GUP reads
\begin{eqnarray}\label{gup1}
\Delta x \Delta p \geq \hbar + \frac{\alpha}{\hbar} \Delta p^{2}.
\end{eqnarray}
 Note that, after solving the quadratic equation for $\Delta p$, we obtain
\begin{eqnarray}\label{gup2}
\Delta p \geq \frac{\hbar}{2 \alpha} \left(\Delta x - \sqrt{\Delta x^{2}-4 \alpha}\right),
\end{eqnarray}
from where
\begin{eqnarray}\label{gup3}
\Delta p \geq \frac{\hbar}{\Delta x} +  \frac{\hbar \alpha}{\Delta x^{3}} + O(\alpha^{2}).
\end{eqnarray}
The above result corresponds to the quadratic minimal length approach. 

\subsection{BH temperature and entropy GUP--corrected}\label{}

In this section, we introduce a generalization of the BH temperature and entropy modified by GUP as reported in Refs. \cite{wen09, xia09, okc19}. To this end, we expand \eqref{gup3} up to second order, namely
\begin{equation}\label{gupml}
\Delta A \geq \Delta x \Delta p \simeq \hbar \left(1 + \frac{\alpha}{\Delta x^{2}} \right) \equiv \hbar'.
\end{equation}
Now, as it is well known, the thermodynamic BH temperature associated with the surface gravity $\kappa$, is given by
\begin{equation}\label{guptemp1}
T = \frac{dA}{dS} \times \frac{\kappa}{8 \pi},
\end{equation}
where
\begin{equation}\label{guptemp2}
\frac{dA}{dS} \simeq \frac{(\Delta A)_{min}}{(\Delta S)_{min}}=\frac{\gamma}{ln2}
\hbar \left(1 + \frac{\alpha}{\Delta x^{2}} \right)
= \frac{\hbar' \gamma}{ln 2}. 
\end{equation}
Then, replacing the relation \eqref{guptemp2} into \eqref{guptemp1}, we obtain
\begin{equation}\label{guptemp3}
T =\frac{\hbar' \gamma}{ln 2}  \times \frac{\kappa}{8 \pi}= \frac{\hbar' \kappa}{2 \pi},
\end{equation}
where we have taken $\gamma=4 ln2$. Finally, after replacing the surface gravity $\kappa$ in the above expression, the GUP--corrected temperature reads 
\begin{eqnarray}\label{guptemp}
T&=&\left(1 + \frac{\alpha}{4 r ^{2}_{+}} \right) \frac{f'(r_{+},P)}{4\pi}\nonumber\\
&=& \left(1 + \frac{\alpha}{4 r ^{2}_{+}} \right) \bigg(2r_{+}P-\frac{h(r_{+},P)}{4\pi r_{+}}-\frac{1}{4\pi}\frac{\partial h(r_{+},P)}{\partial r_{+}} \bigg).
\end{eqnarray}
Note that, in the limit $\alpha\to0$, the expression for the temperature reduces to $T=\frac{\kappa}{2 \pi}$, as expected. As a consequence, the entropy calculated with \eqref{guptemp} acquires a logarithmic correction given by \cite{okc19, mor05}  
\begin{equation}\label{gupentr}
S=\int \frac{dM}{T}=\pi r_{+}^{2} - \frac{\alpha \pi}{4} \log \left( \frac{4 r ^{2}_{+} + \alpha}{\alpha} \right).
\end{equation}

In the following sections, we will use the GUP--corrected temperature \eqref{guptemp} and entropy \eqref{gupentr} within the extended phase space formalism, in order to obtain GUP corrected asymptotically AdS BH solutions.
	
\section{Asymptotically AdS GUP corrected BH Solutions}\label{sec:RD}

\subsection{Minimal length GUP--corrected Polytropic BH}\label{mlgupbhpoly}

\begin{figure}[!h]
    \centering
    \includegraphics*[width=0.5\textwidth]{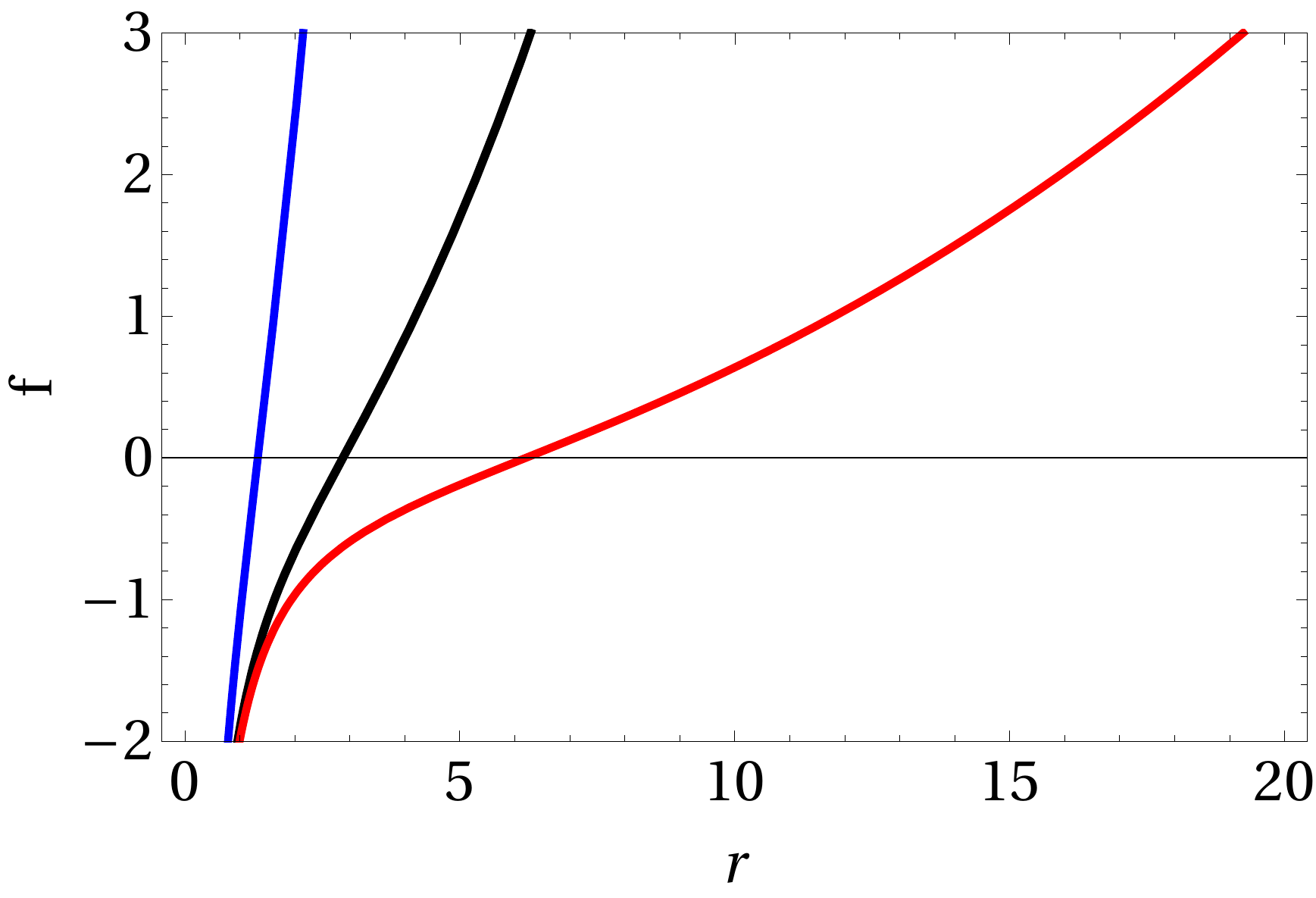}
    \caption[The Polytropic GUP corrected black hole metric function $f$]{$f$ as a function of $r$, with $M=1$, $\alpha=0.1$ and $P=0.1$ (blue line), $P=0.01$ (black line), and $P=0.001$ (red line).}
    \label{fig:1}
\end{figure}

In this section we obtain a GUP corrected polytropic BH solution following the same route taken in section \ref{polysect}. In order to do so, we replace the corrected entropy \eqref{gupentr} and temperature \eqref{guptemp} in the first law \eqref{flt} to obtain a polynomial equation of the form $F_{1}(r)+F_{2}(r)P+F_{3}(r)P^{\frac{1}{1+n}}+F_{4}(r)P^{\frac{n}{1+n}}=0$, where 
\begin{eqnarray}
\label{guppolyf1}F_{1}(r)&=&\frac{C r D(r)}{2(n+1)}-\frac{\alpha +4 r^{2}}{16} \bigg( A'+\frac{A}{r}\bigg).\\
F_{2}(r)&=& \frac{B'(\alpha +4 r^2)}{16}+\frac{B (\alpha -4 r^2)}{16 r}+\frac{\pi r (3 \alpha +4 r^2)}{6}.\\
F_{3}(r)&=&-\frac{1}{16} \bigg(D'(\alpha +4 r^2)+ D \bigg( \frac{4 r (n-1)}{n+1}-\frac{\alpha }{r}\bigg)\bigg).\\
F_{4}(r)&=&-\frac{1}{2} K r B -\frac{4}{3} \pi  K r^3.
\end{eqnarray}
In order to solve the system of equations, lets us impose $F_{4}=0$ to obtain
\begin{eqnarray}\label{guppolyb}
B(r)= -\frac{8 \pi  r^2}{3}.
\end{eqnarray}
Alternatively, we solve for the function $B$, but by setting the condition $F_{2}=0$,
\begin{eqnarray}
B(r)= -\frac{8 \pi  r^2}{3}+\frac{B_{0} (\alpha +4 r^2)}{r}.
\end{eqnarray}
Thus, we conclude that $B_{0}=0$ in order to avoid inconsistencies. Next, by taking the $F_{2}=0$ and solving the differential equation we obtain
\begin{eqnarray}\label{guppolyd}
D(r)=\frac{D_{0} (\alpha +4 r^2)^{\frac{1}{n+1}}}{r}.
\end{eqnarray}
Finally, after imposing $F_{1}=0$ we obtain
\begin{eqnarray}\label{guppolya}
A(r)=\frac{A_{0}}{r}+\frac{D_{0} C (\alpha +4 r^2)^{\frac{1}{n+1}}}{r}.
\end{eqnarray}
Then, replacing \eqref{guppolyb}, \eqref{guppolyd}, and \eqref{guppolya} into the ansatz \eqref{anzatsfltpoly}, the function $h(r,P)$ reads
\begin{eqnarray}\label{guppolyh}
h(r,P)&=& \frac{A_{0}}{r}+\frac{r^2}{l^2}-\frac{3 B_{0} r}{2 \pi l^{2}}\bigg(1+\frac{\alpha}{4 r^{2}} \bigg)+\frac{D_{0} (\alpha +4 r^2)^{\frac{1}{n+1}}}{r} \bigg[\bigg(\frac{3}{8 \pi l^{2}} \bigg)^{\frac{1}{n+1}} + C \bigg].
\end{eqnarray}
Finally, after replacing \eqref{guppolyh} in equation \eqref{f} we arrive to the final result. As a particular case, we set $n=-\frac{1}{3}$, $K=1$, $A_{0}=0$ and $D_{0}=-\frac{\pi  P}{3} (K+P^{\frac{1}{n+1}})^{-1}$ to obtain
\begin{eqnarray}
h = \frac{P 8 \pi r^{2}}{3} - \frac{P \pi(4 r^{2} + \alpha)^{3/2}}{3 r} \label{gph} \\
f = -\frac{2 M}{r}+\frac{P \pi (4 r^2 + \alpha)^{3/2} }{3 r}. \label{guppolyf}
\end{eqnarray}
Note that the AdS structure can be recovered by taking the limit $\alpha \to 0$. In Figure \ref{fig:1} we show the function $f$ for certain parameters. Note that the position of the event horizon increases as the the cosmological pressure decreases.
\begin{figure}[!h]
    \centering
    \includegraphics*[width=0.5\textwidth]{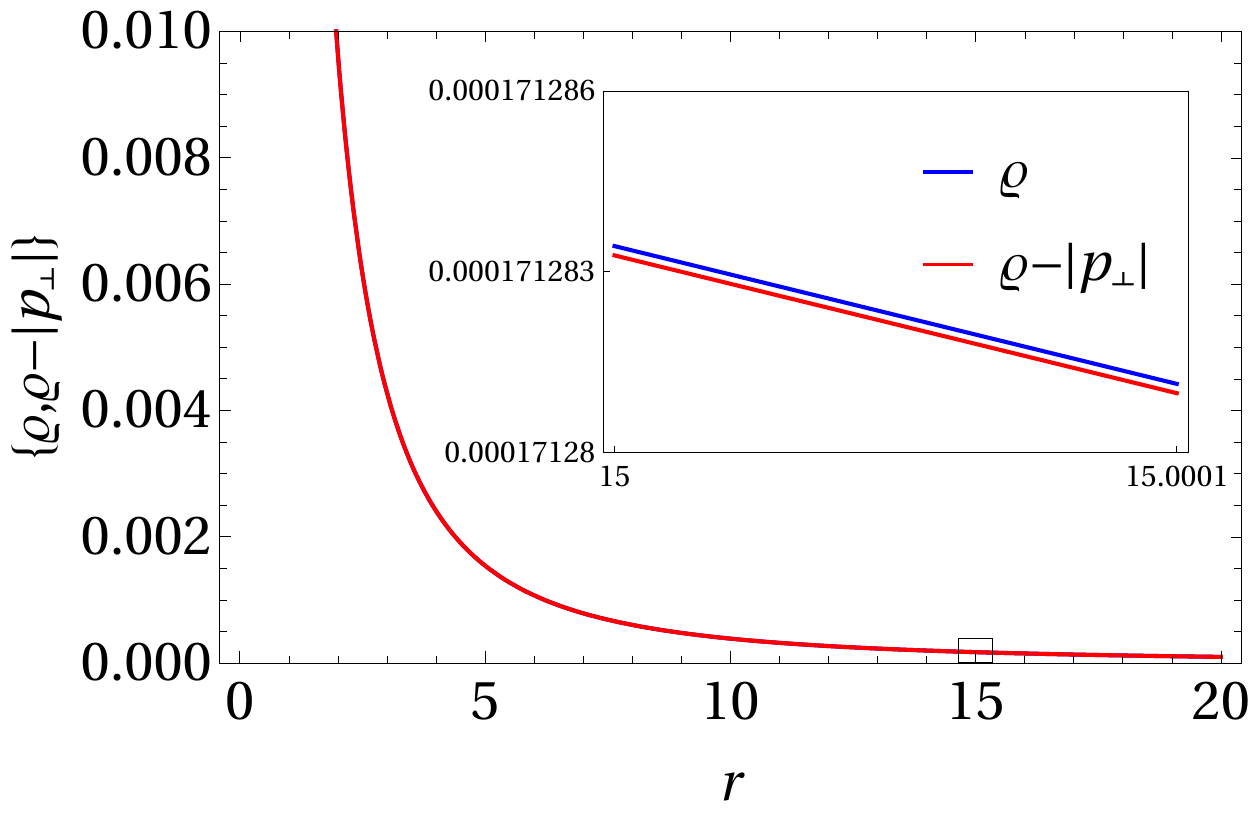}
    \caption[Dominant energy condition of Polytropic GUP corrected black hole]{Dominant energy condition. The energy density $\varrho$ (blue line) and energy density minus the absolute values of perpendicular pressure $\varrho - \lvert p_{\perp} \rvert$ (red line) are plotted by setting $P=0.1$ and $\alpha=0.1$.}
    \label{fig:2}
\end{figure}
\begin{figure}[!h]
    \centering
    \includegraphics*[width=0.5\textwidth]{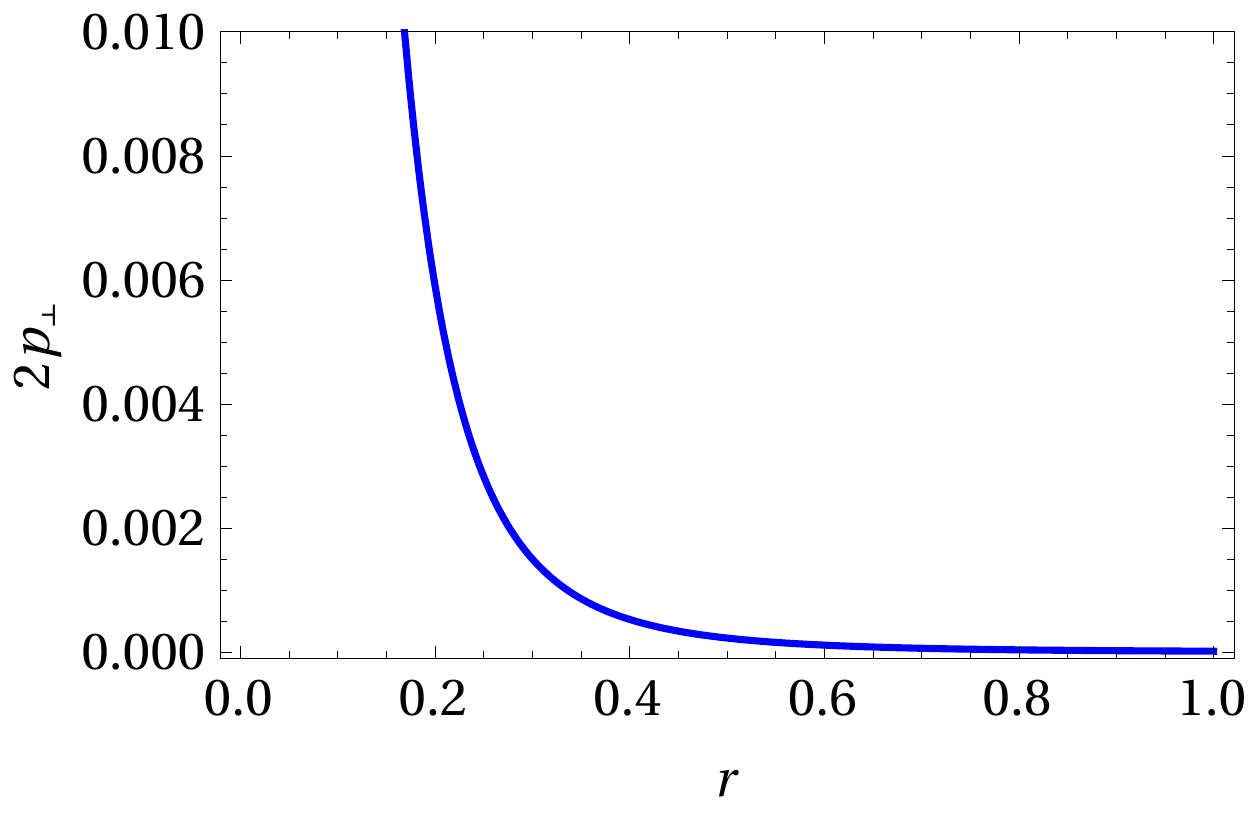}
    \caption[Strong energy condition of Polytropic GUP corrected black hole]{Strong energy condition, by setting $P=0.1$ and $\alpha=0.1$.}
    \label{fig:3}
\end{figure}\\
Using the above results, the matter sector reads
\begin{eqnarray}
\varrho &=& - p_{r} = P+\frac{1}{8 \pi  r^2} -\frac{P \sqrt{\alpha +4 r^2}}{2 r}.\\
p_{\perp} &=& P \left(\frac{\alpha +8 r^2}{4 r \sqrt{\alpha +4 r^2}}-1\right).
\end{eqnarray}
Interestingly, figures \ref{fig:2} and \ref{fig:3} show that the aforementioned solution satisfies both the DEC and the  SEC. In this sense, it is worth noticing that GUP corrections do not alter these energy conditions as in the non-corrected case reported in \cite{set15}.

\subsection{Minimal length GUP--corrected Chaplygin BH}\label{mlgupbhchap}

In this section, we follow the same strategy  implemented in section \ref{chapsect} but this time we assume
\begin{eqnarray}\label{EoSredchap}
P = - \frac{B}{\rho^{n}},
\end{eqnarray}
which corresponds to set $A=0$ in \eqref{EoSchap}. It is worth mentioning that GUP corrections of Chaplygin EoS \eqref{EoSchap} result in very complex differential equations and for this reason, we work with the simplified version.
The resulting polynomial equation takes the form $F_{0}(r)+F_{1}(r)\rho+F_{2}(r)\rho^{-n}+F_{3}(r)\rho^{n+2}=0$, with the corresponding associated functions defined as,
\begin{eqnarray}
F_{0}&=&-\frac{\alpha(r X'+X)(\alpha +4 r^2) }{64 r^3} \bigg[\frac{4 r^2}{\alpha}- \log \bigg(1+\frac{4 r^2}{\alpha}\bigg)\bigg].\\
\label{gupchapf1}F_{1}&=&-\frac{4 \pi r^{3}}{3} -\frac{Y'(\alpha +4 r^2)}{16} - \frac{Y}{16} \bigg[\frac{4 r (n+2)}{n}+\frac{\alpha}{r} \bigg]\nonumber\\
&&- \frac{Z r}{2 B} + \frac{\alpha (r Y'+Y) (\alpha +4 r^2)}{64 r^3} \log \bigg(1+\frac{4 r^2}{\alpha }\bigg).\\
F_{2}&=& -\frac{Z' (\alpha +4 r^2)}{16}  + \frac{B \pi r (3 \alpha +4 r^2)}{6} + \frac{4 r^2-\alpha}{16 r}\nonumber\\
&+& \frac{\alpha (Z'r+Z+B 8 \pi r^{2}) (\alpha +4 r^2)}{64 r^3} \log \bigg(1+\frac{4 r^2}{\alpha }\bigg).\\
F_{3}&=&\frac{r Y}{2 B n}.
\end{eqnarray}
Solving for the function $X$ by setting the state  $F_{0}=0$, we obtain
\begin{eqnarray}\label{gupchapx}
X(r)=\frac{X_{0}}{r}.
\end{eqnarray}
Next, imposing the condition $F_{2}=0$ we arrive at
\begin{eqnarray}\label{gupchapz}
Z(r)=-\frac{8}{3} \pi  B r^2 + Z_{0} \bigg[4 r -\frac{\alpha}{r} \log \bigg(1+\frac{4 r^2}{\alpha}\bigg) \bigg].
\end{eqnarray}
Using  \eqref{gupchapz} in \eqref{gupchapf1}, and replacing the result to solve for the function $Y$ by imposing $F_{1}=0$, we have
\begin{eqnarray}\label{gupchapy}
Y(r)&=&-\frac{n Z_{0}}{B (n+1)} \bigg[4 r -\frac{\alpha}{r} \log \bigg(1+\frac{4 r^2}{\alpha}\bigg) \bigg]\nonumber\\
&&+\frac{Y_{0}}{r} \bigg[4 r^2-\alpha \log \bigg(1+\frac{4 r^2}{\alpha}\bigg)\bigg]^{-\frac{1}{n}}.
\end{eqnarray}
Thus, we can replace \eqref{gupchapx}, \eqref{gupchapz}, and \eqref{gupchapy} in  \eqref{anztasfltchap}, to obtain
\begin{eqnarray}\label{gupchaph}
h(r,P)&=&\frac{X_{0}}{r} + \frac{r^2}{l^{2}} - \frac{3 Z_{0}}{B 8 \pi l^{2}} \bigg[4 r -\frac{\alpha}{r} \log \bigg(1+\frac{4 r^2}{\alpha}\bigg) \bigg]  \nonumber \\
&+& \bigg(-\frac{3}{B 8 \pi l^2}\bigg)^{-\frac{1}{n}} \bigg\{-\frac{n Z_{0}}{B (n+1)} \bigg[4 r -\frac{\alpha}{r} \log \bigg(1+\frac{4 r^2}{\alpha}\bigg) \bigg]\nonumber\\
&+&\frac{Y_{0}}{r} \bigg[4 r^2-\alpha \log \bigg(1+\frac{4 r^2}{\alpha }\bigg)\bigg]^{-\frac{1}{n}}\bigg\}.
\end{eqnarray}
\begin{figure}[!h]
    \centering
    \includegraphics*[width=0.5\textwidth]{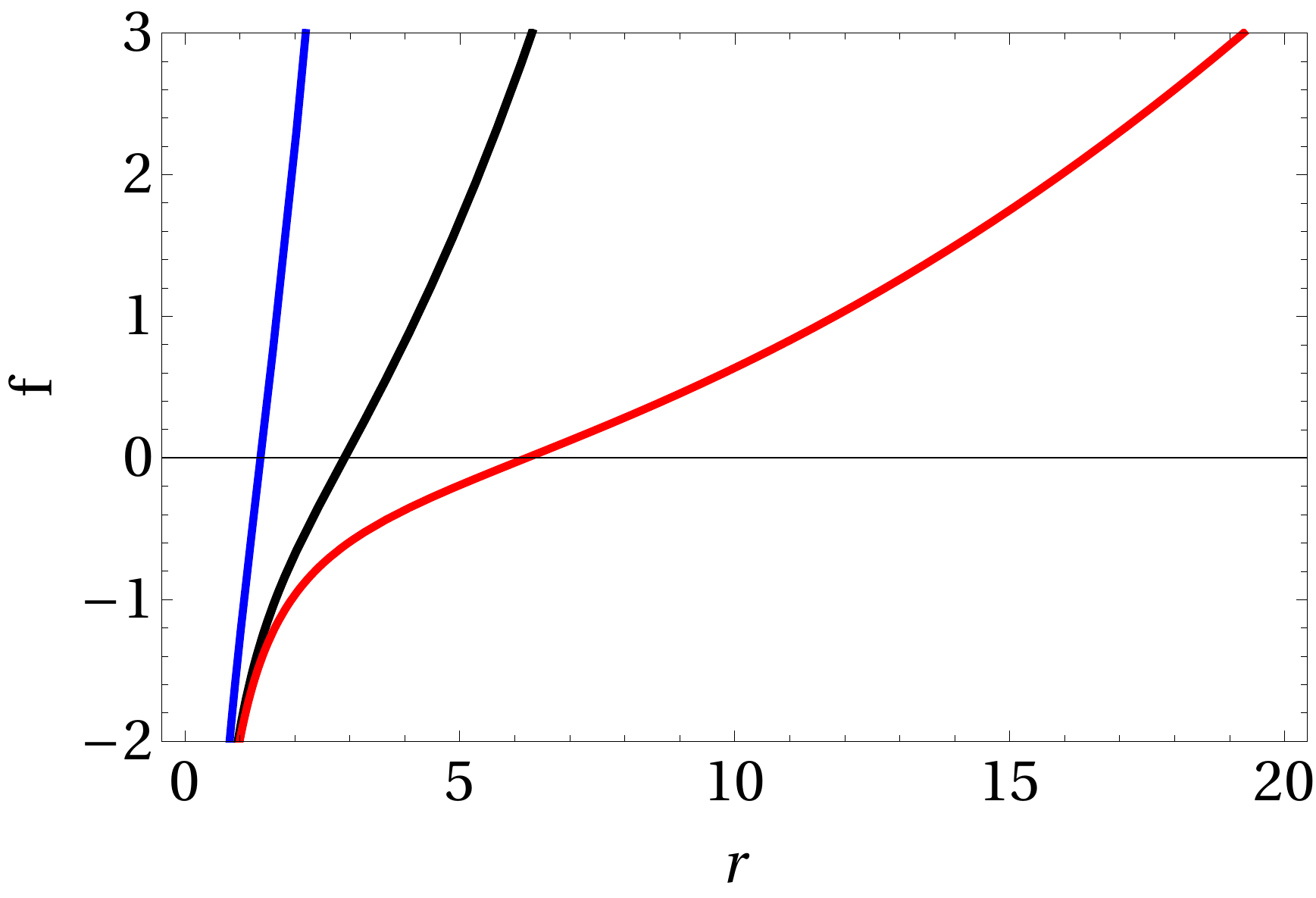}
    \caption[The Chaplygin GUP corrected black hole metric function $f$]{$f$ as a function of $r$, with $M=1$, and $\alpha=0.1$ and $P=0.1$ (blue line), $P=0.01$ (black line), and $P=0.001$ (red line).}
    \label{fig:4}
\end{figure}
\begin{figure}[!h]
    \centering
    \includegraphics*[width=0.5\textwidth]{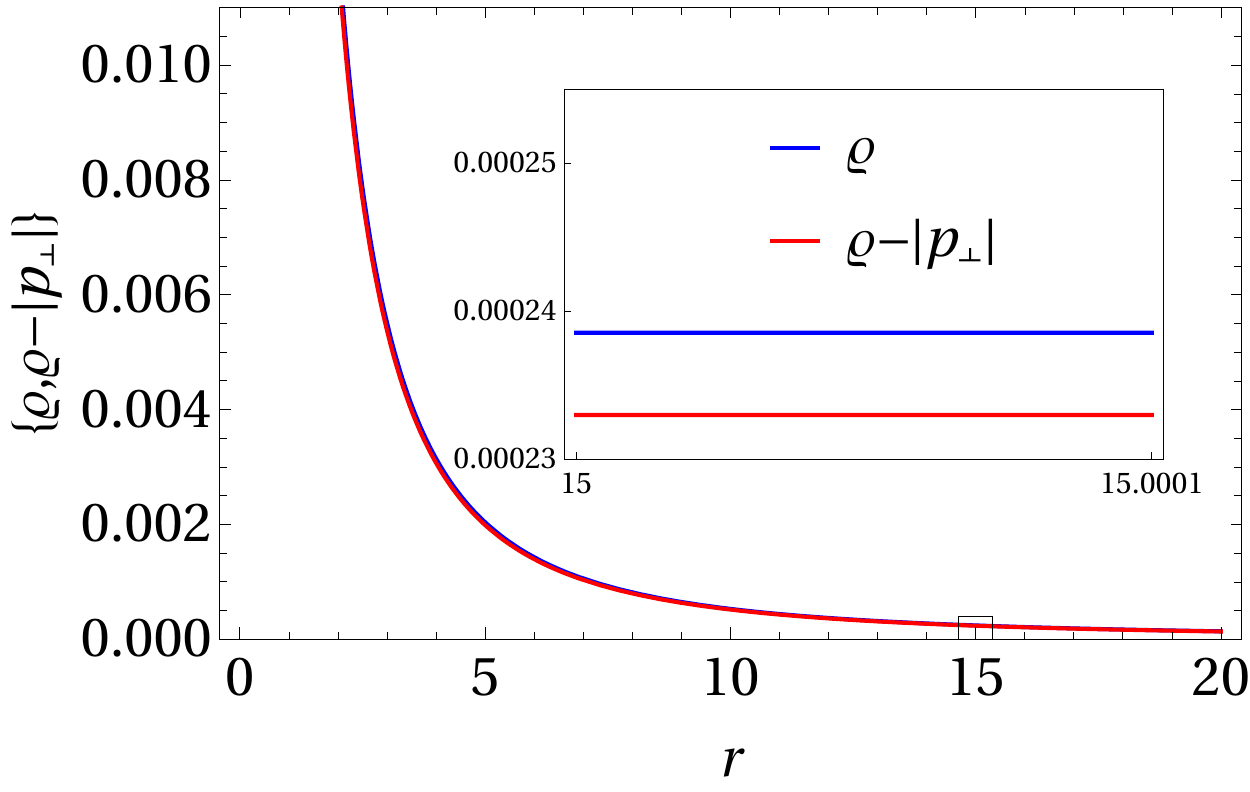}
    \caption[Dominant energy condition of Chaplygin GUP corrected black hole]{Dominant energy condition. The energy density $\varrho$ (blue line) and energy density minus the absolute values of perpendicular pressure $\varrho - \lvert p_{\perp} \rvert$ (red line), by setting $P=0.1$ and $\alpha=0.1$.}
    \label{fig:5}
\end{figure}\\
Finally, we determine the metric function by introducing the solution \eqref{gupchaph} in the equation \eqref{f}. In order to conserve the AdS behavior, we define the values for the constants as $X_{0}=Z_{0}=0$, $Y_{0}=\frac{1}{3} \pi  P \left(-\frac{P}{B}\right)^{\frac{1}{n}}$ and $n=-\frac{2}{3}$, as a result
\begin{eqnarray}
h = \frac{P 8 \pi r^{2}}{3} - \frac{P \pi}{3 r} \bigg[4 r^2-\alpha  \log \left(\frac{4 r^2}{\alpha }+1\right)\bigg]^{3/2} \label{gch} \\
f = -\frac{2 M}{r} + \frac{P \pi}{3 r} \bigg[4 r^2-\alpha  \log \left(\frac{4 r^2}{\alpha }+1\right)\bigg]^{3/2}. \label{gupchapf}
\end{eqnarray}
In this expression, if we take the limit $\alpha \to 0$ then the AdS structure is recovered, as expected. In Figure \ref{fig:4}, we display the solution $f$ as a function of $r$ for some values of $P$.
\\
\\
Finally, the matter sector for this solution reads
\begin{eqnarray}
\varrho &=&  P+\frac{1}{8 \pi  r^2} -\frac{2 P r \sqrt{4 r^2-\alpha  \log \left(\frac{4 r^2}{\alpha }+1\right)}}{\alpha +4 r^2}\\
p_{\perp} &=& -P +\frac{4 P \left(8 r^5+3 \alpha  r^3\right)}{\left(\alpha +4 r^2\right)^2 \sqrt{4 r^2-\alpha \log \left(\frac{4 r^2}{\alpha }+1\right)}}\nonumber\\
&&-\frac{\alpha  P r \left(3 \alpha +4 r^2\right) \log \left(\frac{4 r^2}{\alpha }+1\right)}{\left(\alpha +4 r^2\right)^2 \sqrt{4 r^2-\alpha \log \left(\frac{4 r^2}{\alpha }+1\right)}},
\end{eqnarray}

showing, in Figure \ref{fig:5}, that it satisfies the DEC.

\section{Thermodynamic phase transitions}\label{sec:T}

In this section, we analyze the general thermodynamic properties of the GUP corrected solutions obtained here. Among all the possibilities, we choose the study of the heat capacity phase transitions. Now for reference, we are going to define the procedure in the context of extended phase space without GUP. Then in the following sections, we will promote the obtained results to the GUP versions, and we apply the GUP corrected solutions from previous sections.\\

The heat capacity establishes the thermodynamics stability: a positive heat capacity corresponds to the thermodynamically stable state and a negative heat capacity to the unstable state. If the transition between states is smooth, it corresponds to a first order phase transition. In contrast, a discontinuity in the heat capacity is associated to second order phase transition. The heat capacity of the black hole \cite{sin20, hyu19} is defined as
\begin{eqnarray}\label{hc}
C_{p} = \bigg(\frac{\partial M}{\partial T}\bigg)_{P} = T\bigg(\frac{\partial S}{\partial T}\bigg)_{P},
\end{eqnarray}
at the horizon $r=r_{h}$.
Now, replacing \eqref{temp}  and  \eqref{entr} in \eqref{hc}, we obtain
\begin{eqnarray}\label{heat-cap-}
C_{p} = \frac{2 \pi r^{2}(8 P \pi r^{2} - r h' - h)}{8 P \pi r^{2} + h - r h' - r^{2} h''}\bigg|_{r=r_{h}} \label{HC},
\end{eqnarray}
from where, after setting $h=0$, the heat capacity for an asymptotically AdS BH reads
\begin{eqnarray}
C_{p} = 2 \pi r_{h}^{2} .
\end{eqnarray}

In what follows, we are going to promote equations \eqref{HC}  to their GUP versions to analize the stability of the GUP--corrected asymptotically BH's obtained here.

\subsection{GUP Polytropic BH}
\begin{figure}[!h]
    \centering
    \includegraphics*[width=0.5\textwidth]{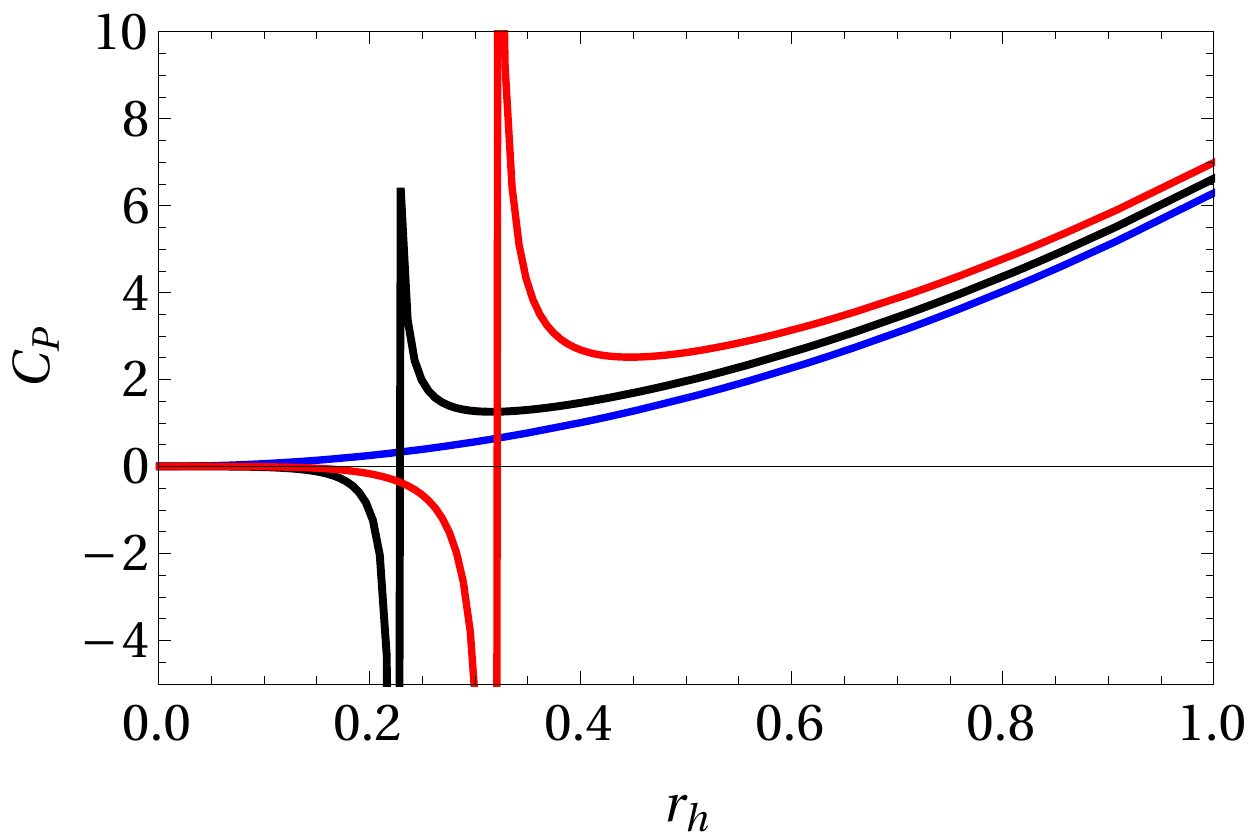}
    \caption[]{The Polytropic GUP corrected Heat Capacity $C_{p}$ as a function of $r_{h}$, the blue line corresponds to $\alpha=0$, the black line to $\alpha=0.1$, and the red line to $\alpha=0.2$.}
    \label{fig:7}
\end{figure}

Following the same strategy introduced in the previous section, the heat capacity for
GUP--correced black holes reads
\begin{eqnarray}
C_{p} = \frac{8 \pi r^{4} (8 P \pi r^{2} - r h' - h)}{(4 r^{2} + 3 \alpha) h + r((4 r^{2}-\alpha)(8 P \pi r - h') - r(4 r^{2}+\alpha) h'')}\bigg|_{r=r_{h}} \label{HC} \label{HCGUP}.
\end{eqnarray}
Note that, Eq. (\ref{heat-cap-}) is recovered in the limit $\alpha \to 0$. Now, replacing
\eqref{gph} in \eqref{HCGUP}, we obtain 
\begin{eqnarray}\label{cppolygup}
C_{p} = \frac{4 \pi r_{h}^{4}}{2 r_{h}^{2} - \alpha}.
\end{eqnarray}
Note that that GUP--parameter in Eq. (\ref{cppolygup}) plays a critical role in the behaviour of the modified $C_{p}$. More precisely, for $\alpha\to0$ the system is stable but when $\alpha>0$ the system undergoes a second order phase transition. In Fig. \ref{fig:7} we show the GUP--corrected $C_{p}$ as a function of the radial coordinate. 

\subsection{GUP Chaplygin BH}

\begin{figure}[!h]
    \centering
    \includegraphics*[width=0.5\textwidth]{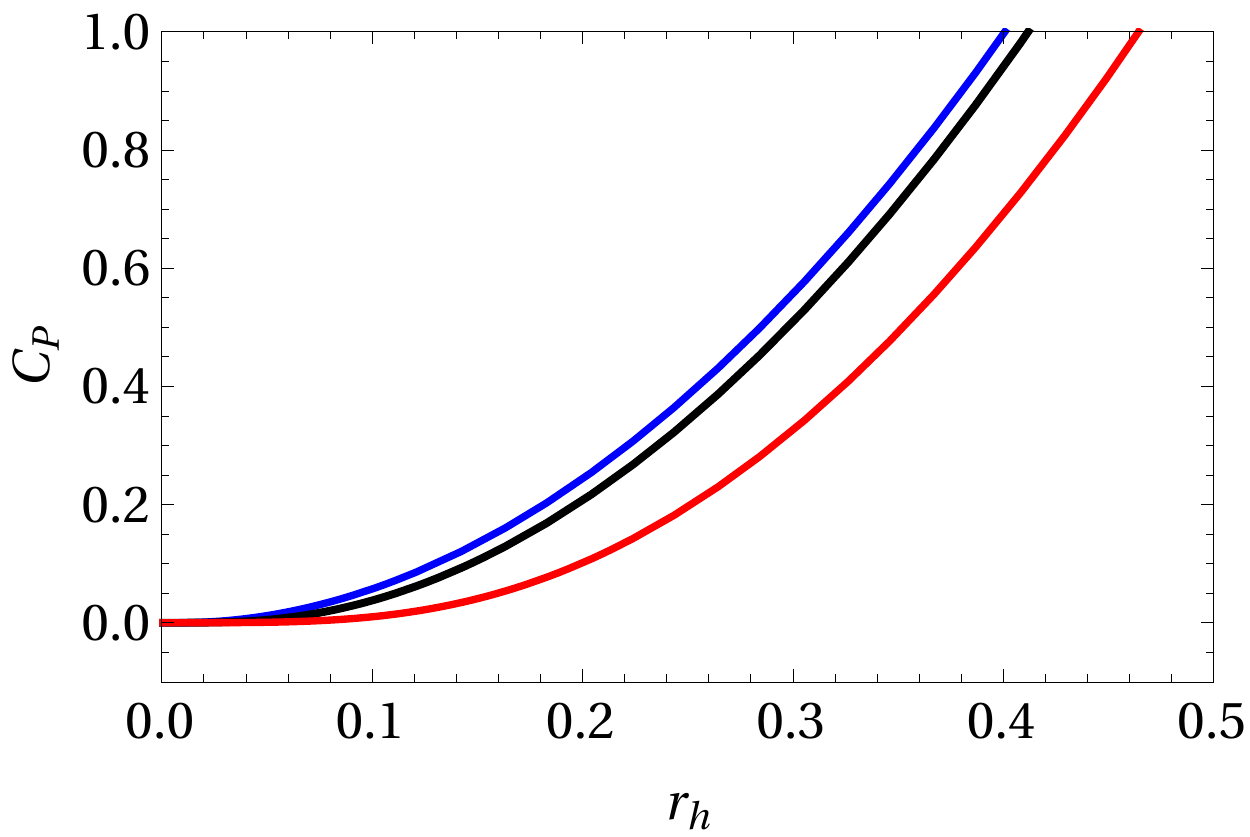}
    \caption[]{The Chaplygin GUP corrected Heat Capacity $C_{p}$ as a function of $r$, the blue line corresponds to $\alpha=0.001$, the black line to $\alpha=0.01$, and the red line to $\alpha=0.1$.}
    \label{fig:9}
\end{figure}

Following the same protocol stated in the previous section of the chapter. We introduce the solution for the GUP Chaplygin BH \eqref{gch} into the GUP corrected heat capacity \eqref{HCGUP} , we obtain 
\begin{eqnarray}
C_{p} = 2 \pi r_{h}^{2} - \frac{\pi \alpha}{2} \log \bigg(1+\frac{4 r_{h}^{2}}{\alpha} \bigg).
\end{eqnarray}
Note that, in contrast to the previous case, the system is stable independent of the value of the GUP parameter.  In figure \ref{fig:9} we plot $C_{p}$ as a function of the radial coordinate for different values of $\alpha$.

\section{Conclusions}\label{sec:CO}
In this work, we obtained asymptotically anti--de Sitter solutions in the context of extended phase space corrected by the generalized uncertainty principle. To be more precise, we used the minimal length corrections provided by the generalized uncertainty principle to modify the black hole temperature and entropy in the first law of the black hole thermodynamics and we close the system by assuming a suitable equation of state for the cosmological pressure which arises in the context of the extended phase space.  In particular, we analyzed the corrections induced on both the Polytropic and Chaplygin black holes. In the Plytropic case, we found that the corrections does not modify the energy conditions fulfilled by the original solution but the system undergoes a second order phase transition for any value of the GUP parameter. In contrast, for the Chaplying case the corrected solution satisfies only the dominant energy condition but its heat capacity is always positive.\\

It is worth mentioning that, on one hand, the solutions we obtained can be thought as semi--classical black holes in the sense that we have incorporated some quantum corrections through the thermodynamics via a GUP-corrected temperature. In this regard, as the GUP leads to universal corrections, it should be interesting if our solutions can be obtained in the framework of other approaches of quantum gravity as Loop Quantum Gravity, for example. On the other hand,
for this work we only review the quadratic minimal length corrections. The combination of other possible corrections could result in unexplored families of BH solutions. This and other features are left for future works.

\subsection*{Acknowledgments}
P. B. is funded by the Beatriz Galindo contract BEAGAL 18/00207 (Spain).

\bibliography{thesis} 
\bibliographystyle{unsrt}
	
\end{document}